# Employment, unemployment and real economic growth

Ivan Kitov
Institute for the Dynamics of the Geopsheres, Russian Academy of Sciences

Oleg Kitov
Department of Economics, University of Oxford

**Abstract**
We have modeled the employment/population ratio in the largest developed countries. Our results show that the evolution of the employment rate since 1970 can be predicted with a high accuracy by a linear dependence on the logarithm of real GDP per capita. All empirical relationships estimated in this study need a structural break somewhere between 1975 and 1995. Such breaks might be caused by revisions to monetary policy (e.g. inflation targeting) or/and changes in measurement units. Statistically, the link between measured and predicted rate of employment is characterized by the coefficient of determination from 0.84 (Australia) to 0.95 (Japan). The model residuals are likely to be associated with measurement errors.

**Introduction**
There is an economic problem which is superior to all other problems in the socio-economic domain. This problem is employment. At all times, paid jobs are the most important source of income. Unemployment is an issue of the same importance but is rather a complementary part of the employment problem. In the short run, the change in unemployment is practically equal to the change in employment, with the level of labour force, i.e. the net change in unemployment and employment, evolving at a lower pace.

It is instructive to use the trade-off between the rate of unemployment and the employment/population ratio to predict the workforce evolution in developed countries. As we have recently revealed, Okun's law (Okun, 1962) is able to provide a very accurate prediction technique for the rate of unemployment (Kitov, 2011). Therefore, one can test real economic growth, as expressed by the change rate of real GDP per capita, as the driving force behind the employment/population ratio. Since the quantitative description of unemployment was very successful with only GDP as a predictor, we do not include in our employment model economic, financial, demographic, educational, or any other variables. Hence, we do not use any of the assumptions underlying the dynamic labour market models introduced and developed by Diamond, Mortensen and Pissarides (e.g. Diamond, 2011; Mortensen and Nagypal, 2007; Pissarides, 2000). Our model is a parsimonious and purely empirical one. We will seek for a theoretical explanation in due course.

We start with a modified Okun's law, where the rate of unemployment is replaced by the employment/population ratio. We also extend the modified Okun's law by integration of both sides of the relevant equation. As a result, one obtains a link between the rate of employment and the logarithm of the overall change in real GDP per capita accompanied by a linear time trend with a positive slope. Thus, our model uses levels instead of differentials.

For the empirical study we use the most recent data on GDP per capita provided by the Conference Board (2011) and data on the employment rate from the U.S. Bureau of Labor Statistics (2011). For several developed countries, the latter time series is available is only from 1970. For the U.S., both variables are available since 1948. We allow for a structural break in the link between the employment rate and real GDP per capita which might manifest artificial changes in definitions of employment and real GDP as well as actual shifts in the linear relationship.

Kitov (2011) has assessed Okun's law in the biggest developed countries: the United States, the United Kingdom, France, Australia, Canada and Spain. The link between employment and real GDP has been studied in the same countries except Spain (BLS does not provide the employment/population ratio). Instead of Spain we included Japan in the list. Our statistical results suggest the presence of a reliable relation between employment and real GDP per capita. The currently low rate of employment is completely related to the lowered rate of real economic growth. As a complementary study, we have calculated empirical coefficients in relevant Okun's law for the U.S.

The remainder of this paper consists of two Sections and Conclusion. Section 1 introduces the integrated version of Okun's law and also presents its modified form for the employment/population ratio. Section 2 describes some empirical results for a number of developed countries.

**1. A modified Okun's law**

According to the gap version of Okun's law, there exists a negative relation between the output gap, $(Y^p-Y)/Y^p$, where $Y^p$ is potential output at full employment and $Y$ is actual output, and the deviation of actual unemployment rate, $u$, from its natural rate, $u^n$. The overall GDP or output includes the change in population as an extensive component which is not necessary dependent on other macroeconomic variables. Econometrically, it is mandatory to use macroeconomic variables of the same origin and dimension. Therefore, we use real GDP per capita, $G$, and rewrite Okun's law in the following form:

$$du = a + bdlnG \qquad (1)$$

where $du$ is the change in the rate of unemployment per unit time (say 1 year); $dlnG=dG/G$ is the relative change rate of real GDP per capita, $a$ and $b$ are empirical coefficients. Okun's law implies $b<0$.

The intuition behind Okun's law is very simple. Everybody may feel that the rate unemployment is likely to rise when real economic growth is very low or negative. An economy needs fewer employees to produce the same or smaller real GDP also because of productivity growth.

When integrated between $t_0$ and $t$, equation (1) can be rewritten in the following form:

$$u_t = u_0 + bln[G_t/G_0] + a(t-t_0) + c \qquad (2)$$

where $u_t$ is the rate of unemployment at time $t$. The intercept $c\equiv0$, as is clear for $t=t_0$. Instead of using the continuous form (2), we calculate a cumulative sum of the annual estimates of $dlnG$ with appropriate initial conditions. By definition, the cumulative sum of the observed $du$'s is the time series of the unemployment rate, $u_t$. Statistically, the use of levels, i.e. $u$ and $G$, instead of their differentials is superior due to suppression of uncorrelated measurement errors.

Figure 1 presents the evolution of an annual increment in the rate of unemployment in the U.S. and that of the employment/population ratio, $e$, with a negative sign, $-de$. Both curves evolve practically in sync from 1949 to 2010 with the $du$ series demonstrating a slightly higher volatility. A linear regression gives for the curves in Figure 1 a slope of 1.24 and $R^2=0.88$. It is an important observation that $du$ almost coincides with $-de$ in amplitude after 1985. The high level of correlation between these variables allows for a modified Okun's law for $de$:

$$de = \alpha + \beta dlnG \qquad (3)$$

where $\beta$ should be a positive constant. Equation (3) is the sought functional dependence between the first differences of the rate of employment and the growth rate of real GDP per capita.

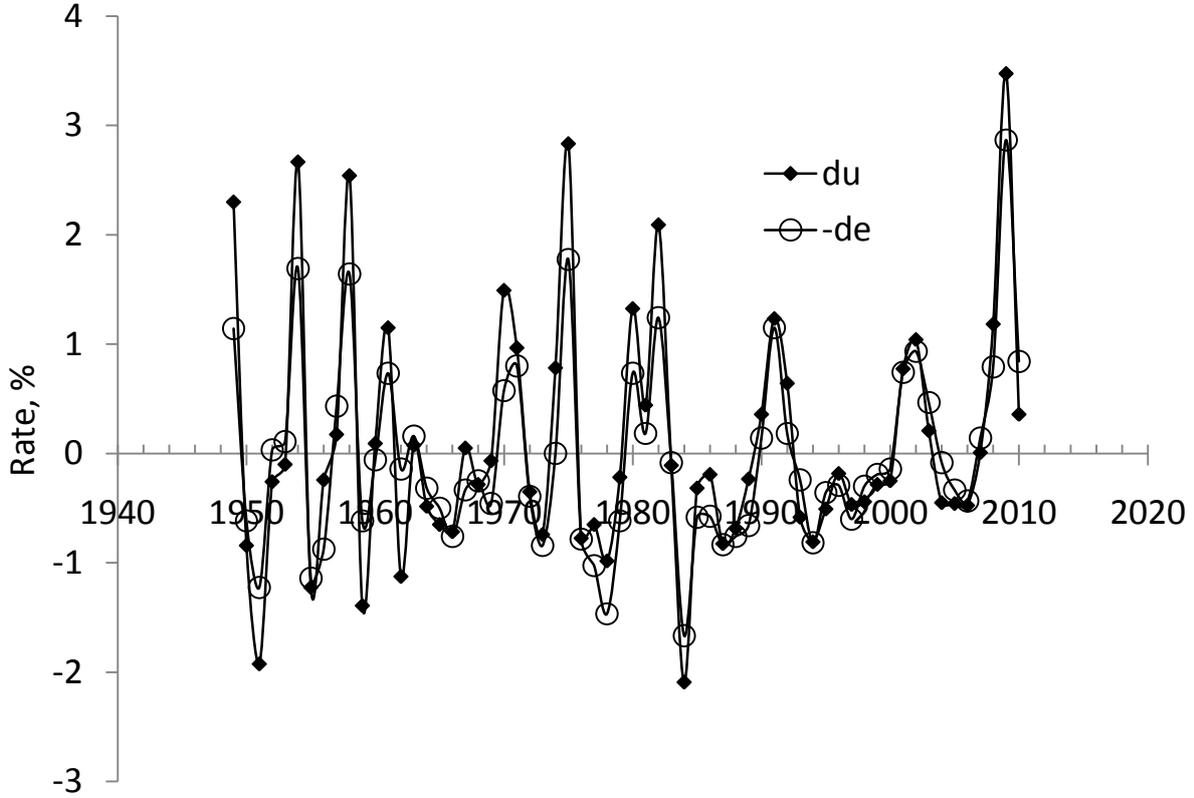

**Figure 1. The evolution of *du* and *–de* in the U.S. between 1949 and 2010.**

When integrated, relationship (3) has the following form:

$$e_t = e_0 + \beta ln[G_t/G_0] + \alpha(t-t_0) \tag{4}$$

In this study, we use (4) for the estimation of both coefficients by a standard LSQ method. We seek for the best fit between the measured and predicted rate of employment.

Kitov (2011) showed the necessity of a structural break in (1). Figure 1 also suggests that the link between *de* and *dlnG* might change around 1985 because the relationship between *du* and *de* also changes. Therefore, we introduce a floating structural break in (4) which year has to be fixed by the best fit as well. For the U.S., we allow the break year to vary between 1975 and 1995. Thus, relationship (4) should be split into two segments:

$$e_t = e_0 + \beta_1 ln[G_t/G_0] + \alpha_1(t-t_0), \ t<t_s$$
$$e_t = e_s + \beta_2 ln[G_t/G_{ts}] + \alpha_2(t-t_s), \ t \geq t_s \tag{5}$$

where $e_0$ is the measured employment/population ratio at time $t_0$; $e_s$ is the predicted employment/population ratio at the time of structural break $t_s$; $\alpha_1$ and $\beta_1$, $\alpha_2$ and $\beta_2$ are empirical coefficients estimated before and after the structural break, respectively. The integral form of Okun's law should be also split into two time segments:

$$u_t = u_0 + b_1 ln[G_t/G_0] + a_1(t-t_0), \ t<t_s$$
$$u_t = u_s + b_2 ln[G_t/G_{ts}] + a_2(t-t_s), \ t \geq t_s \tag{6}$$

## 2. Empirical results

We start with the case of U.S. Since we have already estimated a preliminary empirical relationship for the rate of unemployment using only the best visual fit (Kitov, 2011), it is important to re-estimate Okun's law using a more reliable statistical procedure. The method of least squares applied to the integral form of Okun's law (6) results in the following relationship:

$$du_p = -0.406dlnG + 1.113, \quad 1979 > t \geq 1951$$
$$du_p = -0.465dlnG + 0.866, \quad 2010 \geq t \geq 1979 \tag{7}$$

where $du_p$ is the predicted annual increment in the rate of unemployment, $dlnG$ is the relative change rate of real GDP per capita per one year. There is a structural break around 1979 which divides the whole 60-year interval into two practically equal segments. Figure 2 displays the measured and predicted rate of unemployment in the U.S. since 1951. The agreement between these curves is excellent with a standard error of 0.55%. The average rate of unemployment for the same period is 5.75% with the mean annual increment of 1.1%. All in all, this is a very accurate model of unemployment with $R^2=0.89$. Hence, the model (the integral Okun's law) explains 89% of the variability in the rate of unemployment between 1951 and 2010 with the model residual likely related to measurement errors. Statistically, there is no room for other variables to influence the rate of unemployment, except they are affecting the real GDP per capita.

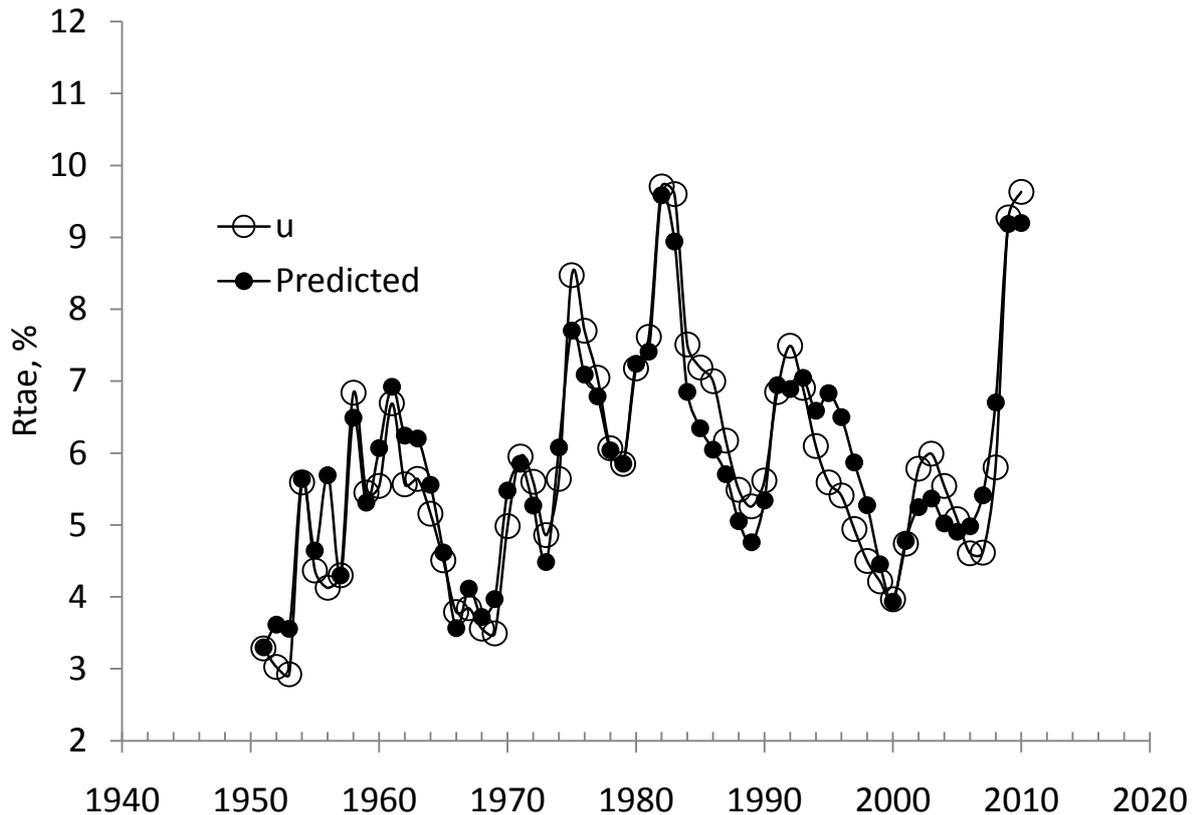

**Figure 2. The observed and predicted rate of unemployment in the USA between 1951 and 2010.**

In (7), the rate of real GDP growth has a threshold of (0.866/0.465=) 1.86% per year for the rate of unemployment to be constant. When $dlnG$ is larger than this threshold the rate of unemployment in the U.S. starts to decrease. Figure 3 displays the evolution of $dlnG$ since 1979. On average, the rate of growth was 1.65% per year, i.e. slightly lower than the threshold and the rate of unemployment has been increasing since 1979.

Empirical model (7) suggests a tangible shift in slope and a significant change in intercept around 1979. This is a very important finding. There are two terms in (7) which define the evolution of the unemployment rate: real economic growth counteracts the positive linear time trend. Figure 4 depicts both components. The difference or the distance between $a(t-t_0)$ and $-b\ln(G_t/G_0)-u_0$ in Figure 4 is the rate of unemployment.

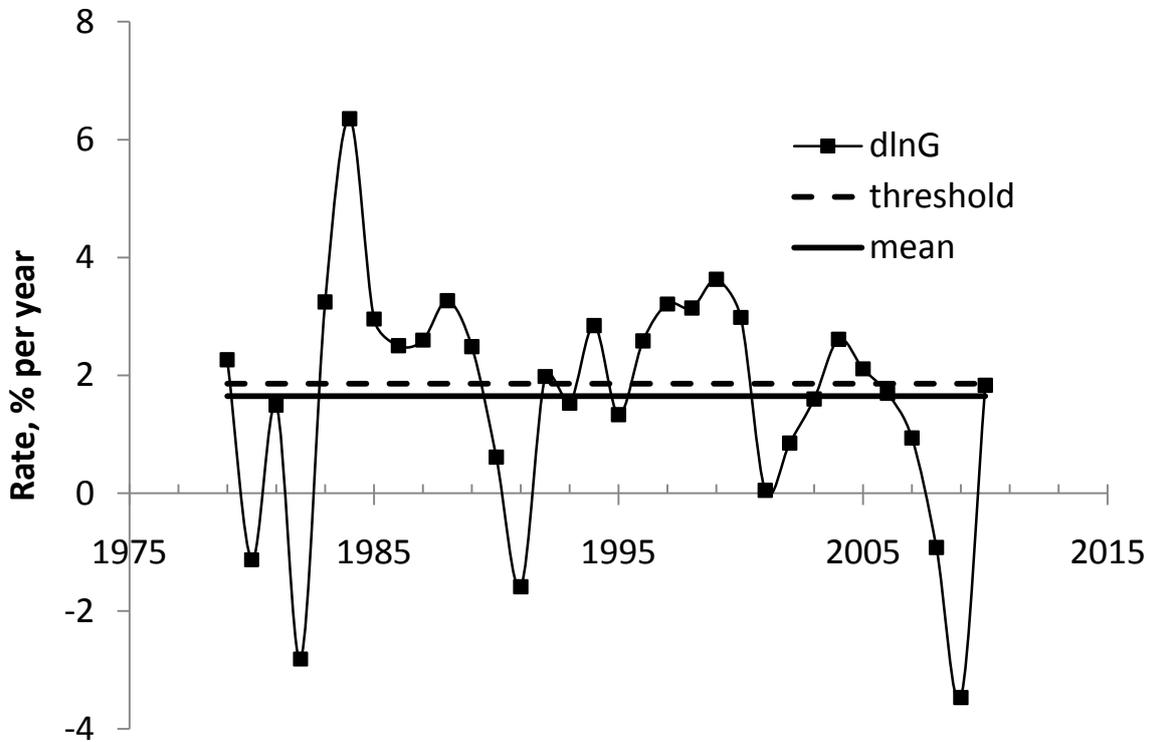

**Figure 3.** *dlnG* as a function of time. Also shown is the threshold 1.86% per year and the mean growth rate 1.65% per year.

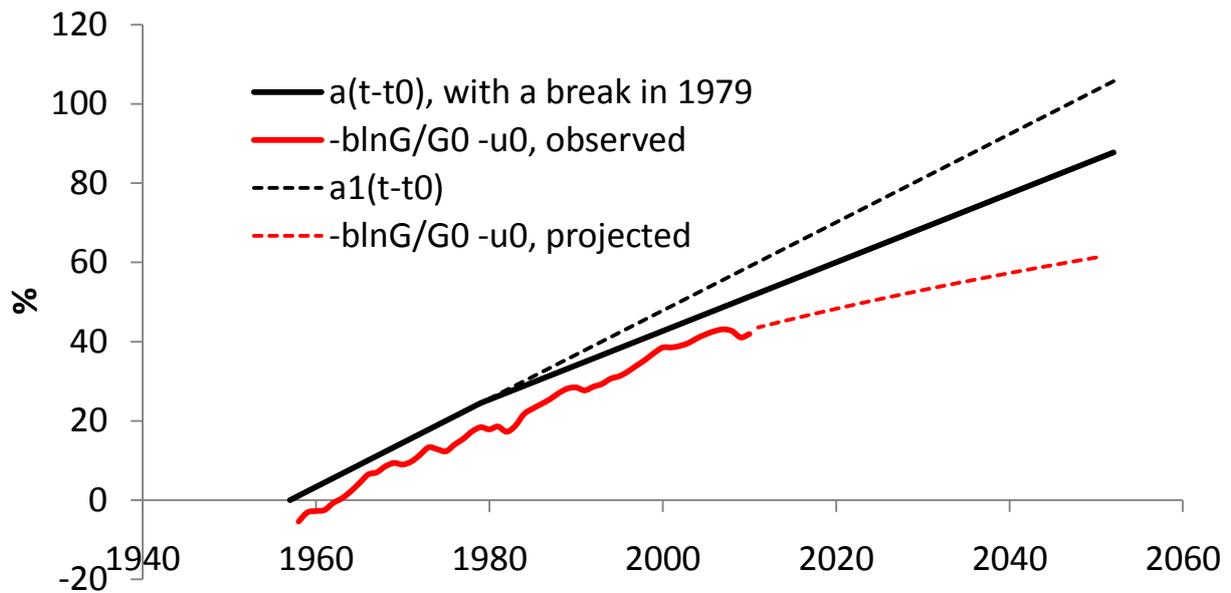

**Figure 4.** The evolution of two components in (7) defining the unemployment rate.

The importance of the break in 1979 is obvious when we extend the trend $a_1(t-t_0)$ observed before 1979. The distance would have been much larger with the old trend after 1979, i.e. the rate of unemployment would have been also higher than that actually measured. If to extrapolate the current time trend and the dependence on *G* one can project the rate of unemployment. Figure 4 also depicts such a projection through 2050. Without a new structural break, the rate of unemployment in 2050 will be near 25%. It might happen that the U.S. is currently struggling through a transition to a new relation in (7) which will be able to keep the rate of unemployment below 10%. In any case, the growth rate of real GDP per capita has to be much higher than 2% per year in order to reduce the current rate of unemployment to the level of 5%.

In the above paragraph, we used a constant increment of real GDP per capita to extrapolate its evolution into the future. For a constant annual increment, one obtains a logarithmic time trend which follows from our model of economic growth (Kitov, 2009). We have introduced an inertial growth component, $(dlnG/dt)_i$, which is inversely proportional to the attained level of real GDP per capita:

$$(dlnG/dt)_i = C/G \qquad (8)$$

where *C* is an empirically estimated constant. The term *C/G* represents the inertial rate of growth. Figure 5 demonstrates the observed evolution of *G* since 1950 and gives two projections: a linear one with an annual increment *C*=$591.5 and an exponential growth following the trend observed before 2010. The deviation between these projections is fast and the next few years should help to distinguish between them.

In the long run, the evolution of $G_t$ is linear over time: $G_t=G_0+C(t-t_0)$. Then two terms in (6) have different time trends (logarithmic and linear, respectively) and $u_t$ must grow with time if there are no structural breaks.

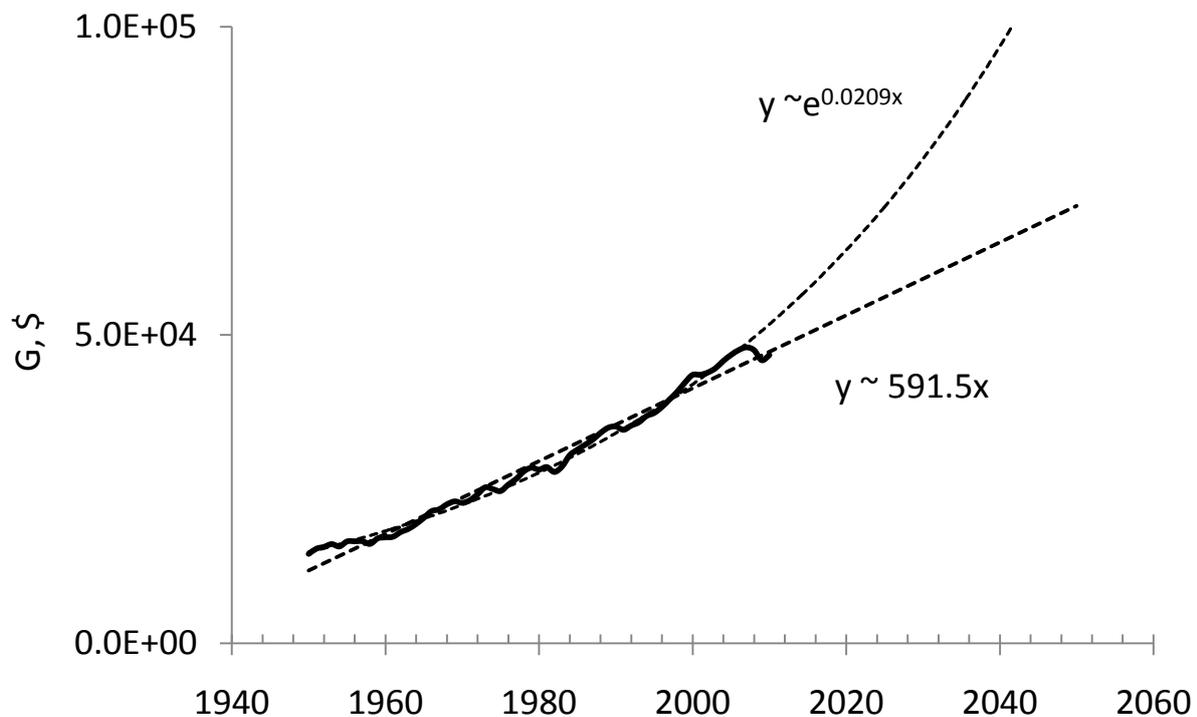

**Figure 5. The evolution of *G* over time with a projected linear trajectory for *C*=$591.5 and an exponential trajectory $G_t=G_0exp(0.0209t)$, where the exponent corresponds to that obtained for the period between 1950 and 2010.**

After obtaining relationship (7) with a high predictive power for the rate of unemployment we have estimated (using the integral form) a modified Okun's law for the employment/population ratio:

$$de = 0.277dlnG – 0.457, \quad 1983>t\geq1951$$
$$de = 0.496dlnG – 0.870, \quad 2010>t\geq1983 \qquad (9)$$

Figure 6 compares the observed and predicted change in the employment/population ratio. Figure 7 shows the cumulative curves for both time series in Figure 6. It also provides some hints on the nature of the break near 1983. (The difference between 1979 and 1983 as the estimated break years might be related to the effect of measurement noise on the results of the least squares method.) The employment/population ratio jumped from ~57% in 1982 and ~63% in 1989. The change in slope in (7) and (9) is rather similar: both the rate of employment and unemployment is more sensitive to the rate of change in GDP after the break.

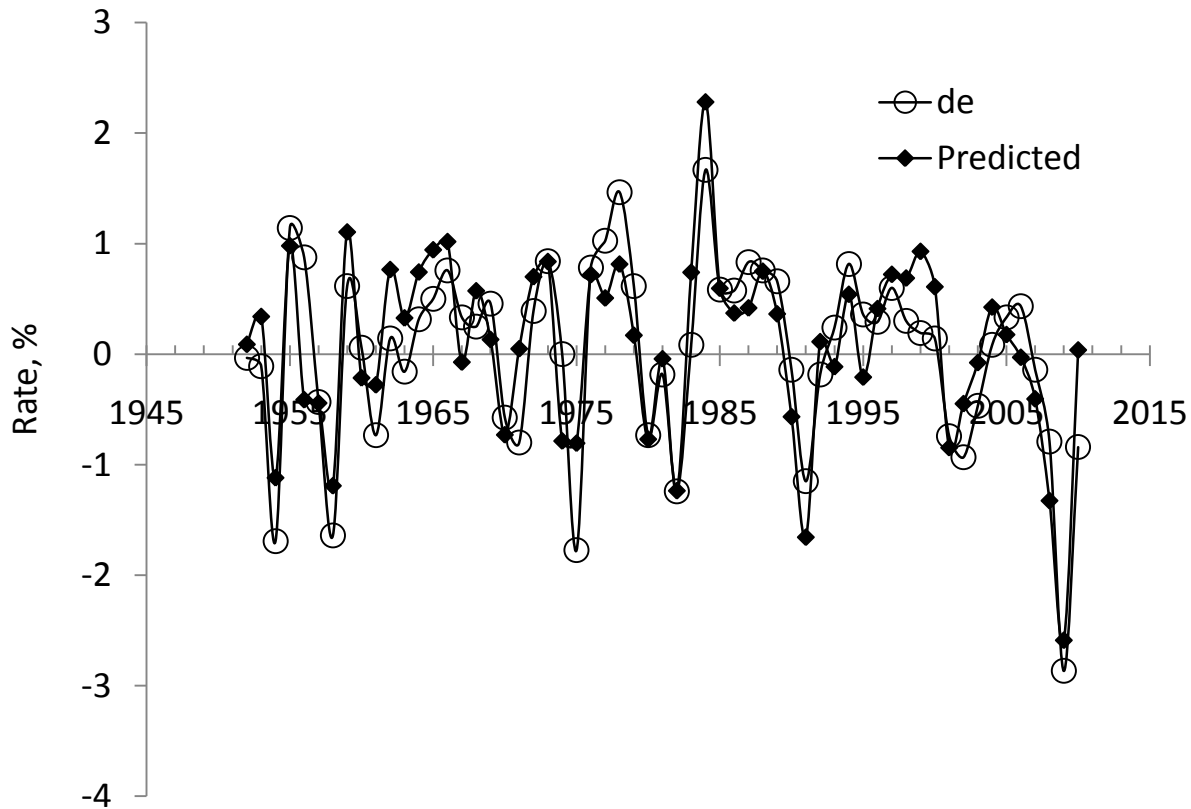

**Figure 6. The observed and predicted change in the employment/population ratio, *de*.**

The next country to model is the United Kingdom. The change in the unemployment rate is also highly correlated with the change in employment/population ratio, $R^2=0.79$. Figure 8 shows both curves between 1972 (the employment rate estimates are available since 1971) and 2010. As expected, the change in the rate of unemployment is more volatile.

The best fit model for the employment/population ratio in the UK is as follows:

$$de_t = 0.41dlnG_{t-1} – 1.11, \quad 1983>t>1971$$
$$de_t = 0.41dlnG_{t-1} – 0.81, \quad 2010\geq t\geq1983 \qquad (10)$$

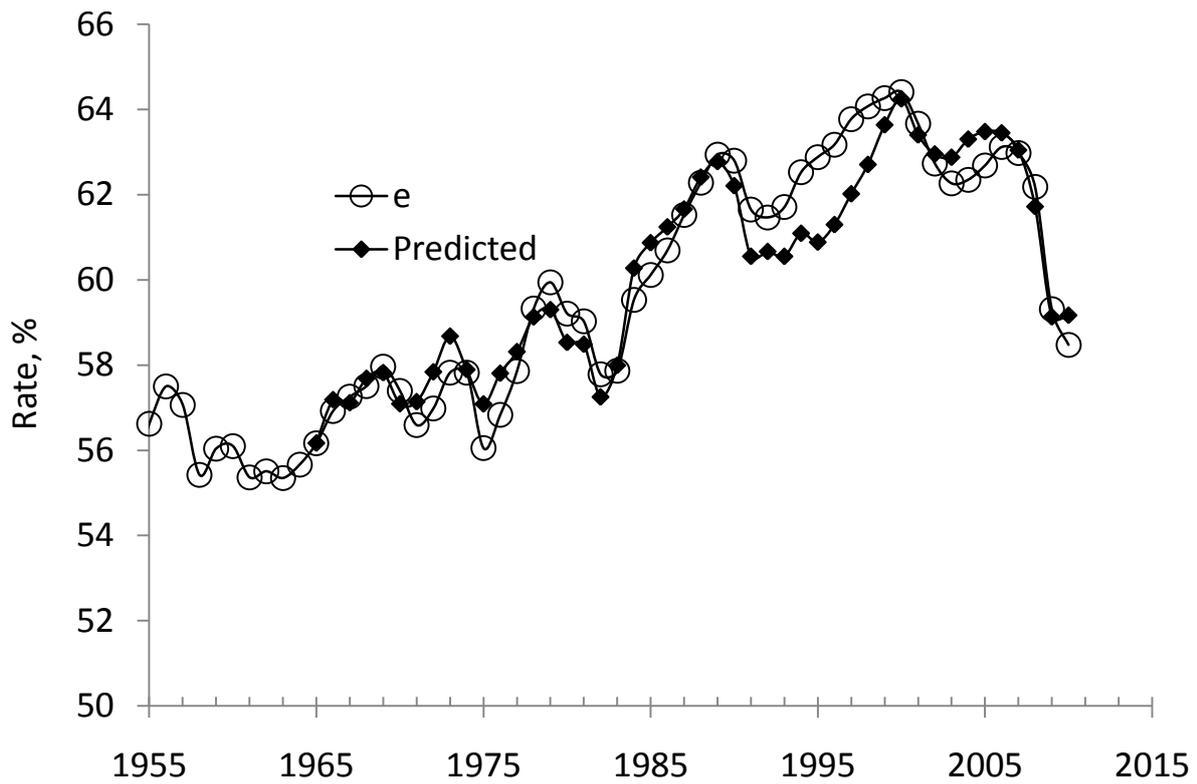

**Figure 7. The cumulative curves for the observed and predicted change in the employment/ population ratio in the U.S.**

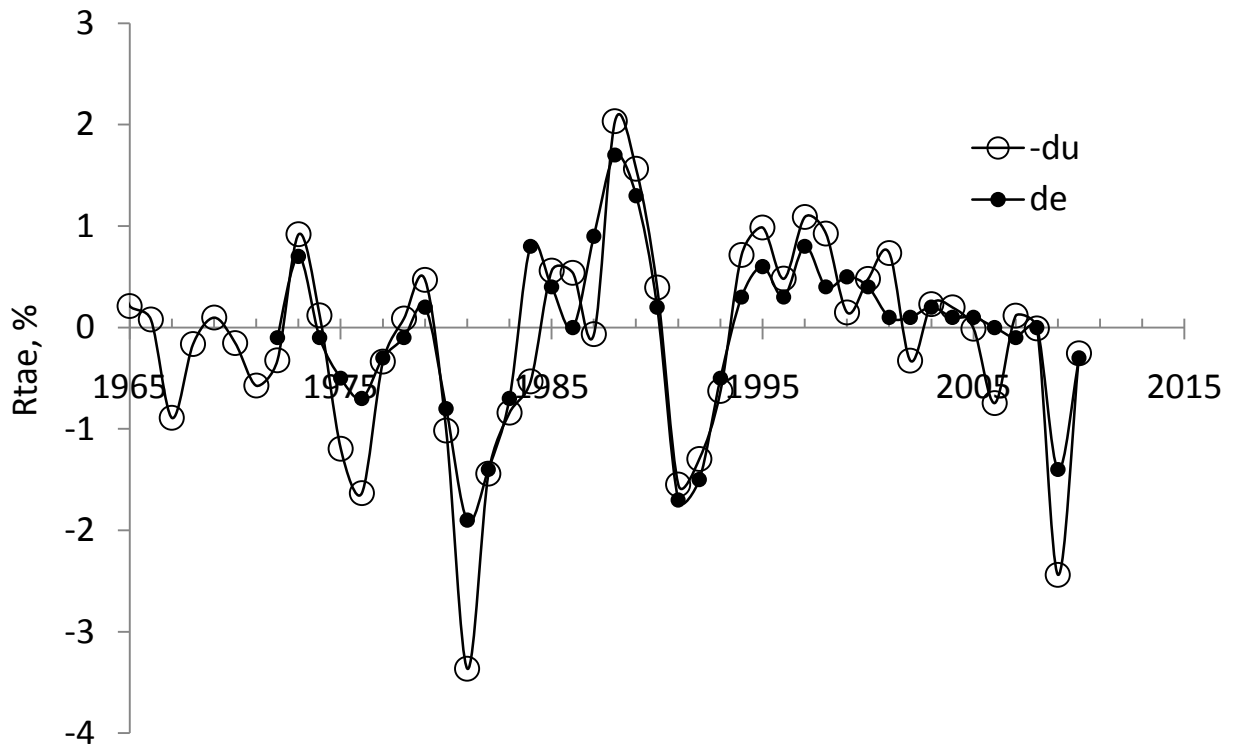

**Figure 8. The (negative) change in the rate of unemployment compared to the change in the rate of employment in the UK.**

where $dlnG_{t-1}$ is the change rate of real GDP per capita one year before, i.e. the predicted curve leads by one year. Figure 9 shows the cumulative curves for the time series in (10). The agreement between the curves is excellent with $R^2=0.89$. The standard error is 0.47% between 1972 and 2009. There is a large deviation from the measured employment/population ratio after 2008. This deviation is consistent with the difference between *de* and *–du* in 2009.

There is a structural break near 1983 which is expressed by a significant shift in intercept without any change in slope. The employment/population ratio varies around 58% between from ~54.3% in 1982 and ~61% in 1972. For the period after 1983, relationship (10) implies that the UK needs the rate of GDP growth above (0.81/0.41=) 1.97% per year to increase the employment rate. However, the discrepancy between the measured and predicted curves after 2008 may manifest the start of a transition to a new dependence in (10), with a lower sensitivity of the employment rate to the change in GDP.

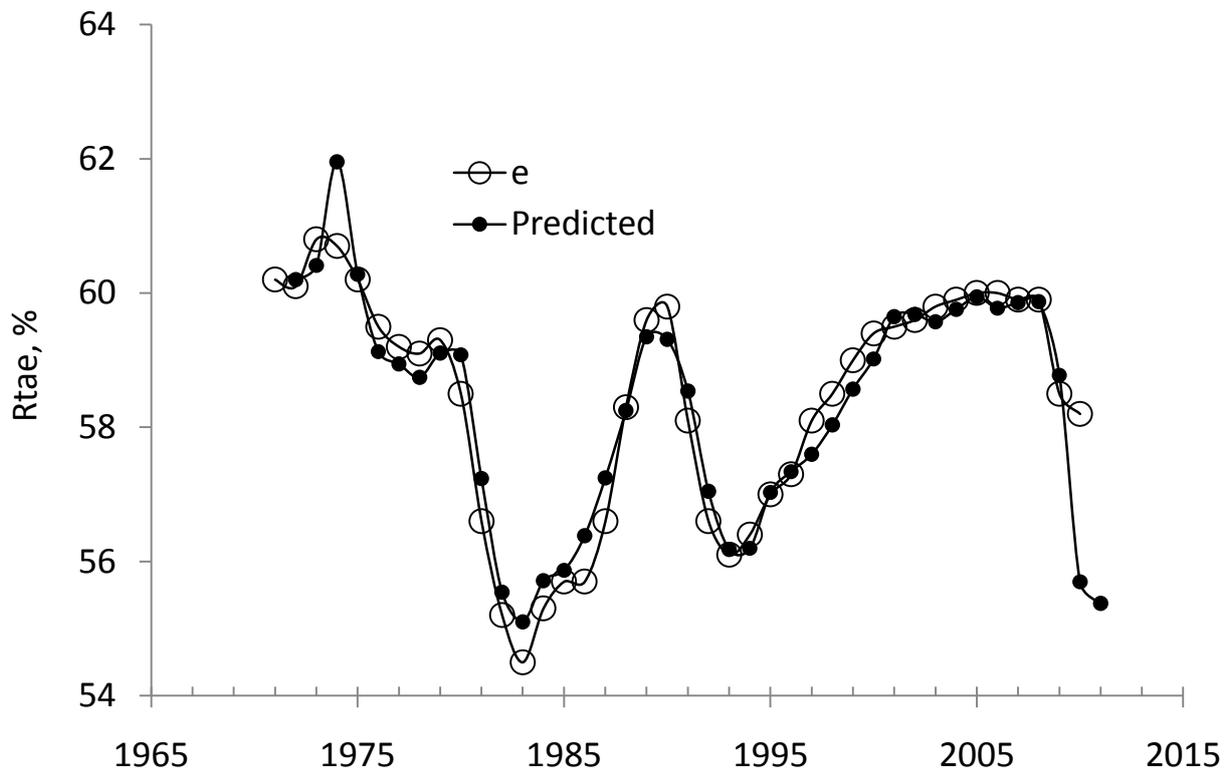

**Figure 9. The cumulative curves for the observed and predicted change in the employment/ population ratio in the UK.**

In Canada, the change in the rate of employment, *de*, and the rate of unemployment, *du*, are well correlated ($R^2=0.76$) as Figure 10 demonstrates. As in other developed countries, the rate of unemployment is more volatile. For Canada, the following best-fit model has been obtained by the least-squares (applied to the cumulative sums):

$$de_t = 0.40dlnG_t - 0.67, \ 1984>t>1970$$
$$de_t = 0.44dlnG_t - 0.56, \ 2010 \geq t \geq 1984 \tag{11}$$

where $dlnG_t$ is the change rate of real GDP per capita at time *t*. Figure 11 shows the cumulative curves for the time series in (11). The overall fit is very good with $R^2=0.84$ and a standard error of 0.83% between 1971 and 2010. Before 1984, the predicted curve is likely leading the measured one by 1 year. It is important that the most recent period is well described and the fall in the employment rate in 2009 was completely driven by the drop in real GDP per capita.

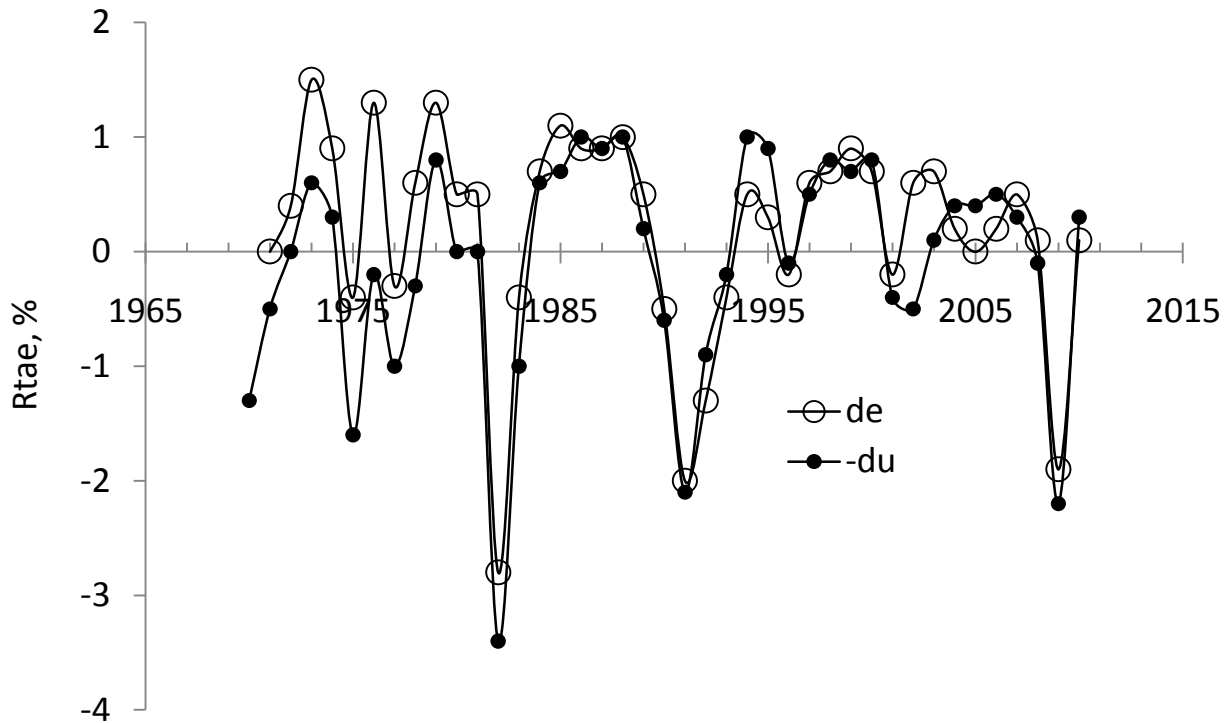

**Figure 10. The (negative) change in the rate of unemployment compared to the change in the rate of employment in Canada.**

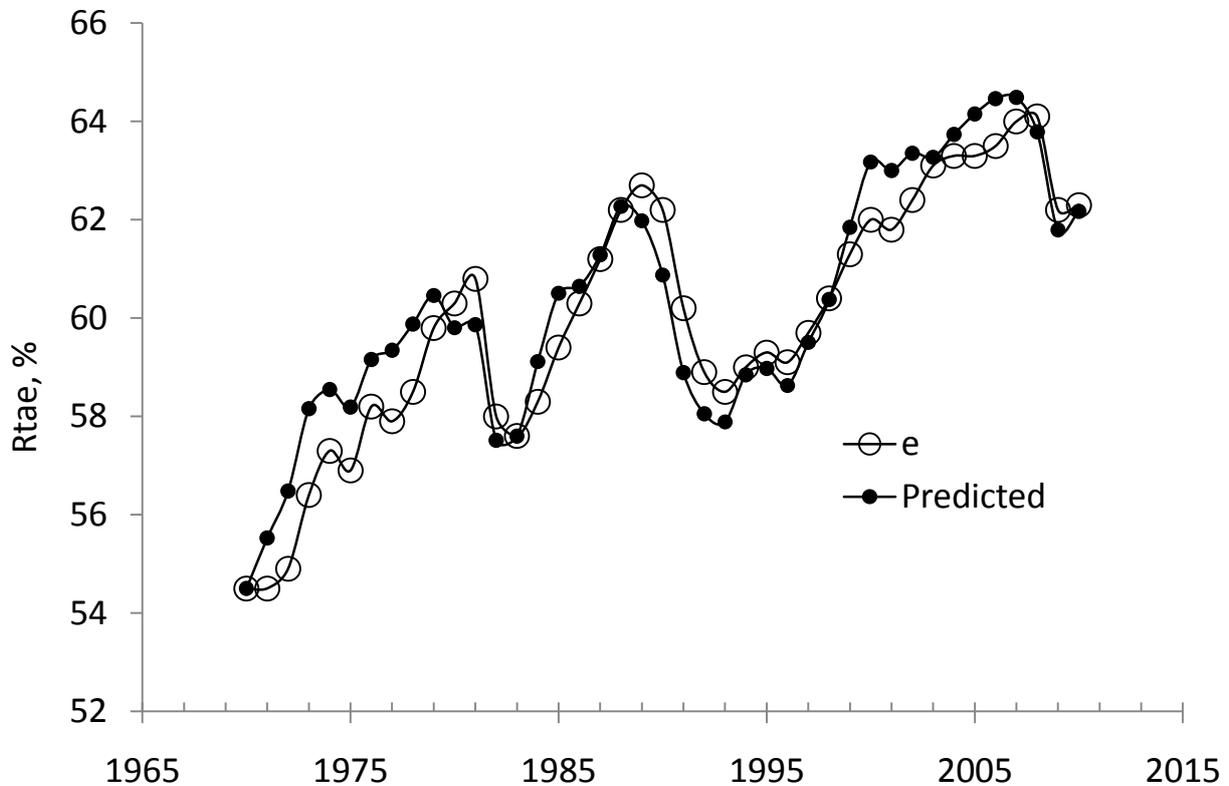

**Figure 11. The cumulative curves for the observed and predicted change in the employment/ population ratio in Canada.**

There is a structural break near 1984 which is expressed by a slight shift in slope and intercept. The employment/population ratio varies between from ~54.5% in 1971 and ~64.1% in 2008. For the period after 1984, relationship (11) demands the rate of GDP growth above (0.56/0.44=) 1.27% per year for the employment rate to increase. Otherwise, the employment rate will be falling and the rate of unemployment will be increasing.

France gives another example of successful modeling. As expected, the change in the rate of unemployment in Figure 12 is more volatile except the shift in the employment rate near 1982. This is a completely artificial break from 53.2% in 1981 to 55.3% in 1982, and we do not need to model it. It is worth noting that $-du$ and $de$ diverge in 2010. This means that Okun's law for France does not fit observations.

For France, the best-fit model for the rate of employment is as follows:

$de_t = 0.155dlnG_t – 0.65$, $1994>t>1970$
$de_t = 0.250dlnG_t – 0.30$, $2010 \geq t \geq 1994$     (12)

Figure 13 shows the cumulative curves for the time series in (12). There is a structural break near 1994 which is expressed by significant shifts in slope and intercept. The employment/population ratio varies between from ~56%% in 1970 and 50.4% in 1992. The agreement is good with $R^2=0.91$ and a standard error of 0.39% for the period between 1971 and 2010. The rate of employment will grow when the rate of GDP growth exceeds (0.30/0.25=) 1.2% per year.

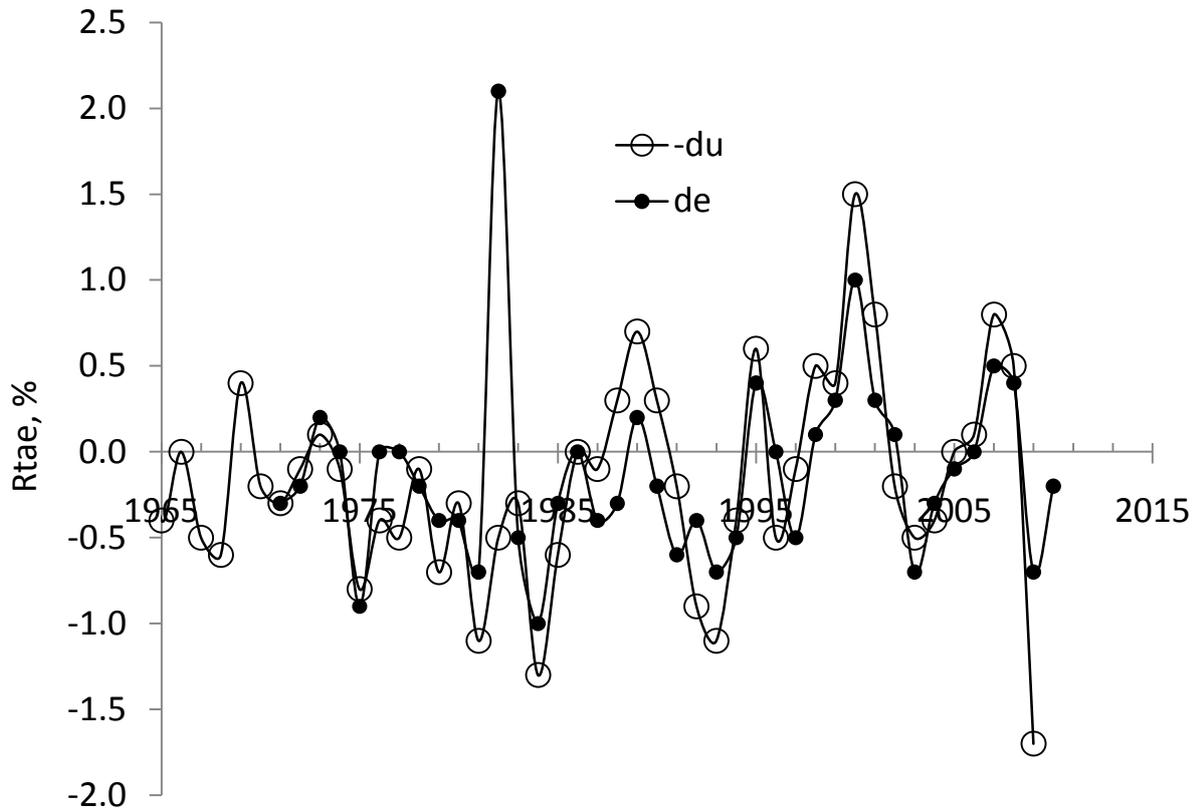

**Figure 12. The (negative) change in the rate of unemployment compared to the change in the rate of employment in France.**

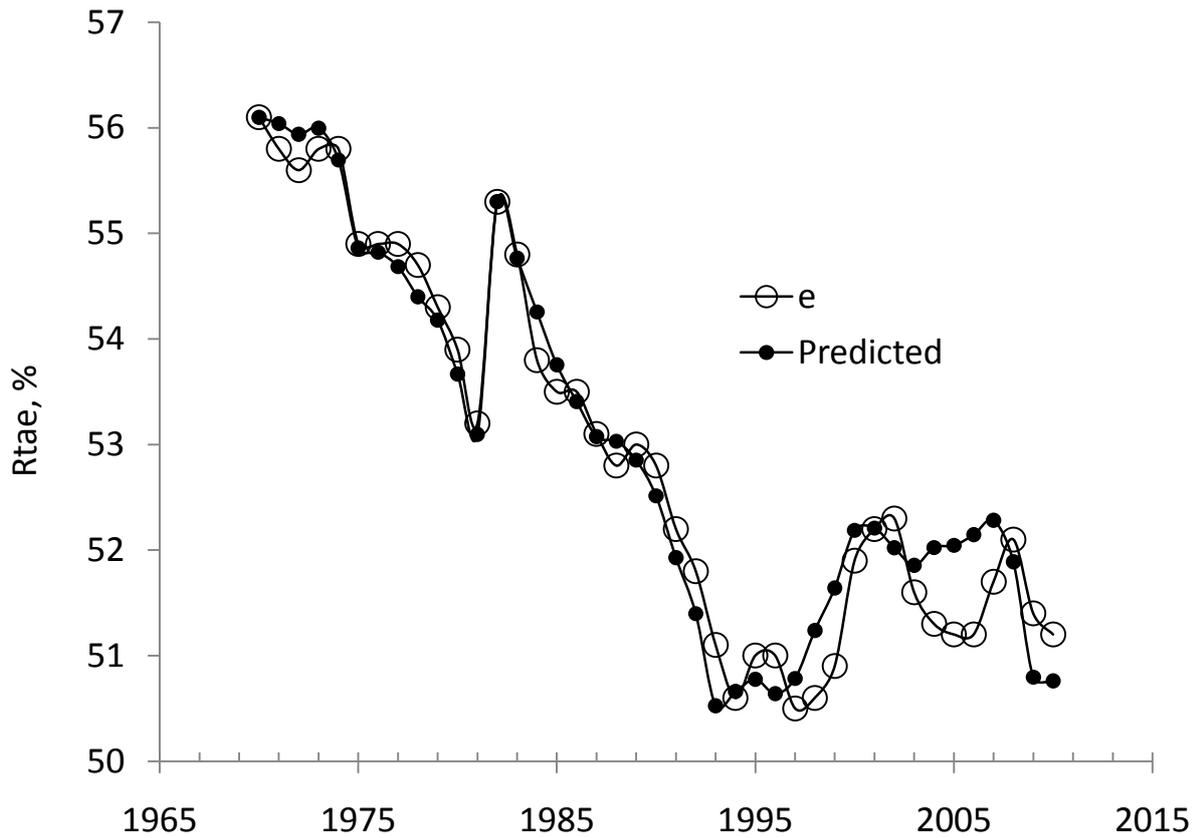

**Figure 13. The cumulative curves for the observed and predicted change in the employment/population ratio in France.**

Figure 14 compares the change in the rate of employment and the rate of unemployment, in Australia. The best-fit model for the rate of employment is as follows:

$de_t = 0.50dlnG_t – 0.92$, $1983>t>1970$
$de_t = 0.41dlnG_t – 1.08$, $2010\geq t\geq 1983$ (13)

Figure 15 shows the cumulative curves for the time series in (13). There is a structural break near 1983 which is expressed by tangible shifts in slope and intercept. The employment/population ratio varies between from 55%% in 1983 and 64% in 2008. The agreement is very good with $R^2=0.84$ and a standard error 1.19% for the period between 1971 and 2010.

We finish modeling the evolution of the employment rate in developed countries with Japan. Figure 16 compares the change in the rate of employment, *de*, and the negative rate of unemployment, *-du*, in Japan. The latter variable is as volatile as former one and they differ drastically ($R^2=0.45$) compared to the synchronized evolution of these variables in the U.S. That's why we have failed to obtain a reasonable Okun's law for Japan.

The employment/GDP model for Japan is similar to Okun's law. The best-fit model has been obtained by the least-squares (applied to the cumulative sums):

$de_t = 0.02dlnG_t – 0.53$, $1978>t>1950$
$de_t = 0.14dlnG_t – 0.42$, $2010\geq t\geq 1978$ (14)

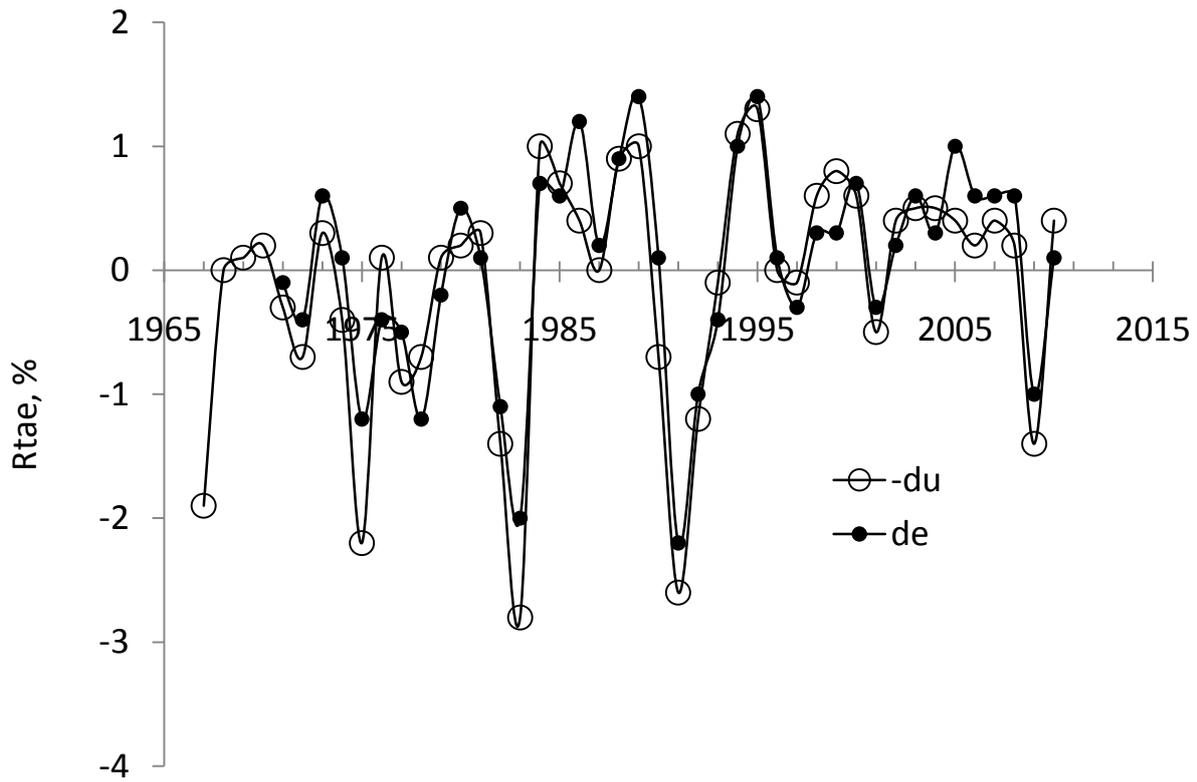

**Figure 14. The (negative) change in the rate of unemployment compared to the change in the rate of employment in Australia.**

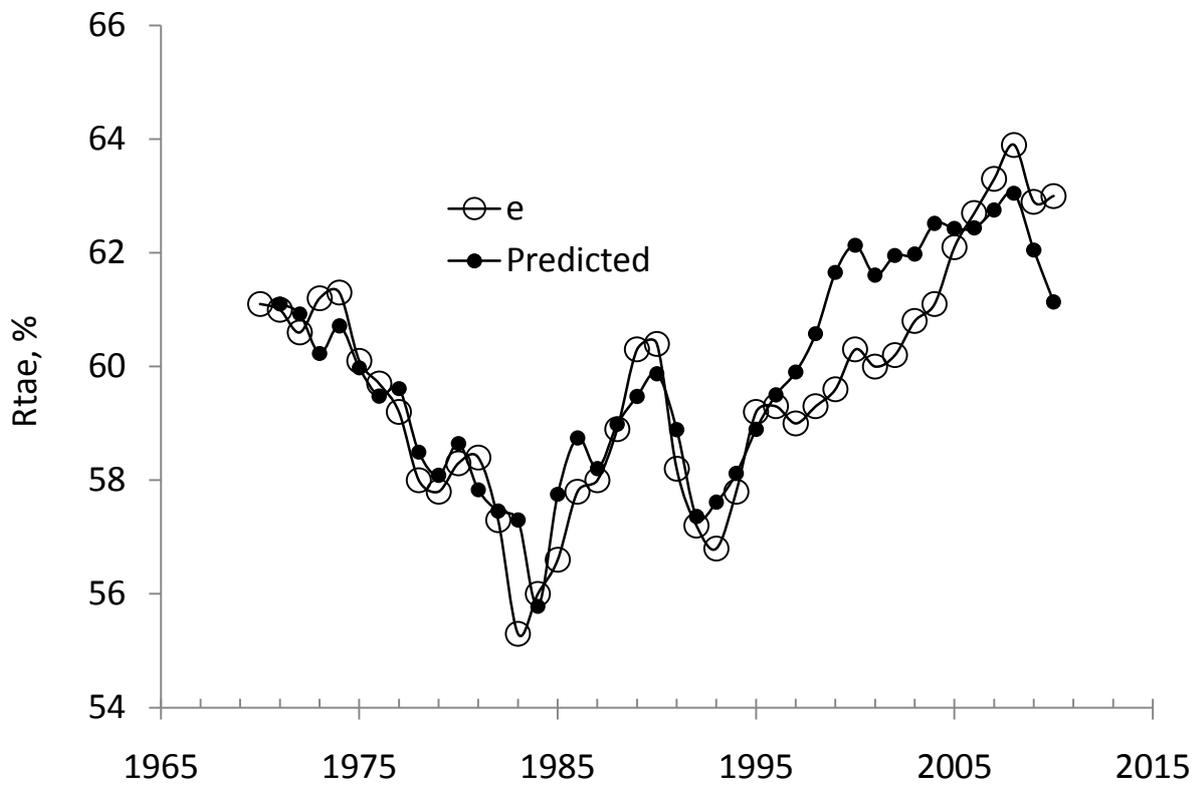

**Figure 15. The cumulative curves for the observed and predicted change in the employment/ population ratio in Australia.**

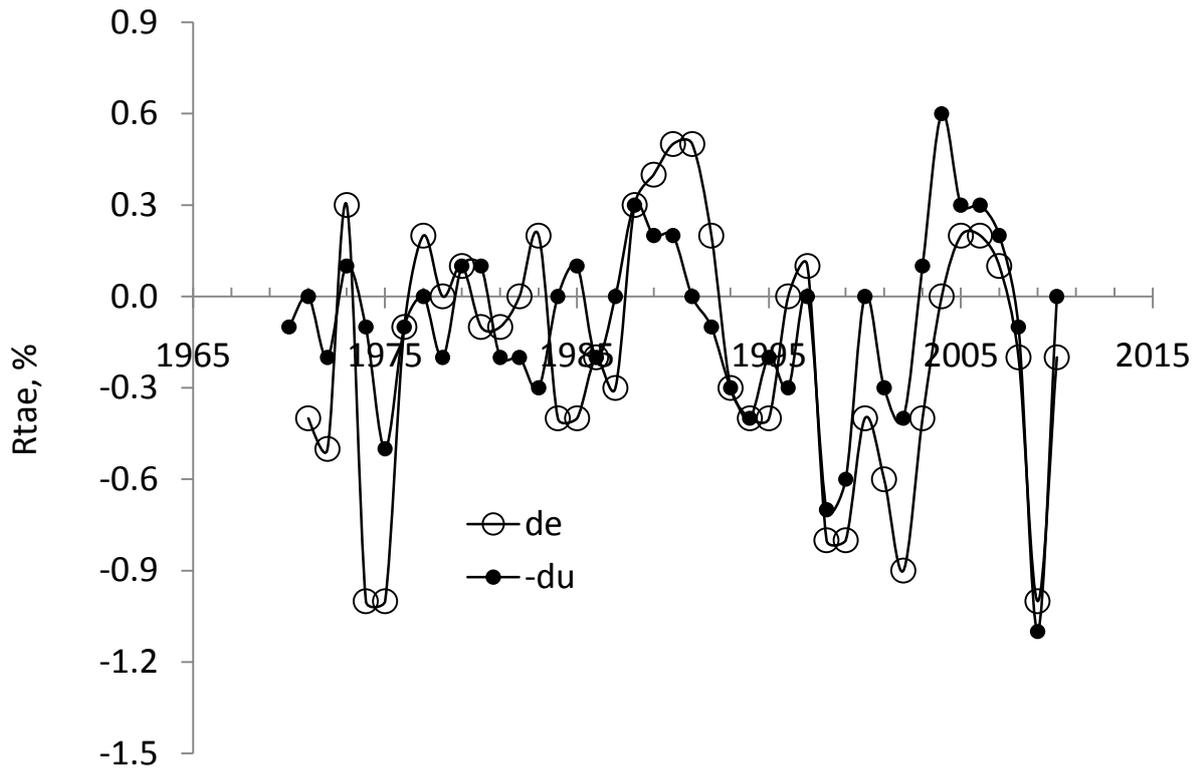

**Figure 16.** The (negative) change in the rate of unemployment compared to the change in the rate of employment in Japan.

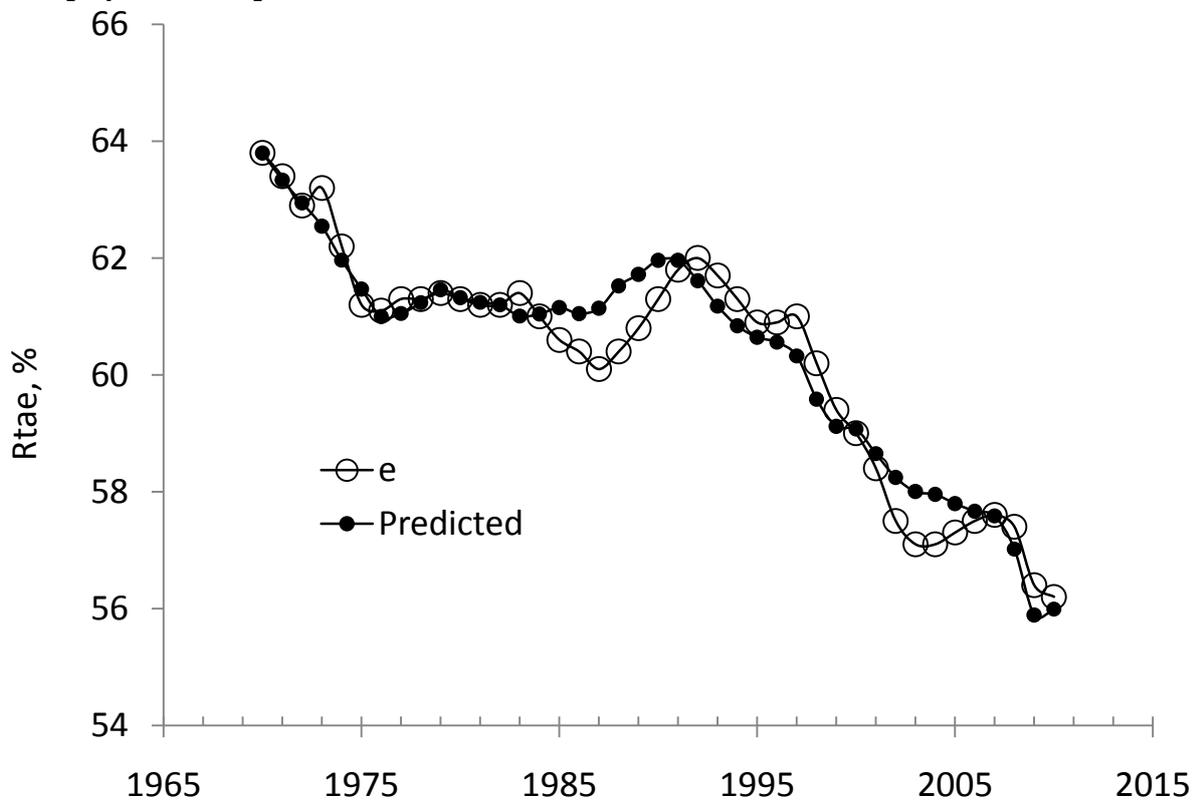

**Figure 17.** The cumulative curves for the observed and predicted change in the employment/ population ratio in Japan.

Figure 17 shows the cumulative curves for the time series in (14). There is a structural break in 1978 which is expressed by a dramatic shift in slope and a slight break in intercept. The employment/population ratio varies between from 64%% in 1970 and 56% in 2010. The agreement is excellent with $R^2$=0.95 and a standard error 0.50% for the period between 1971 and 2010.  The coefficient of determination might be biased up when both time series are nonstationary. In the long run, we consider both variables as stationary ones despite the negative trend since 1970.

**Conclusion**

We have modeled the evolution of employment/population ratio in the biggest developed countries using a modified Okun's law with the rate of change of real GDP per capita as the driving force. This model demonstrates an extraordinary predictive power with the coefficient of determination between 0.84 and 0.95. One can accurately describe the dynamics of employment (and thus, unemployment) since 1970.